\documentstyle[twoside,11pt,draft]{article}
\def\p-{$\pi^-\,$}

\def\het{$^3\!H\!e\,$}
\def\he4{$^4\!H\!e\,$}

\begin{document}
\input epsf
%
\par
\vspace{1cm}
\begin{center}
{\Large \bf Origin of light nuclei in near earth orbit}  
\vspace{0.3cm}
\par
          L.Derome and M. Bu\'enerd
\par
\vspace{0.2cm}
{\sl Institut des Sciences Nucl\'{e}aires, IN2P3, 
53 av. des Martyrs, 38026 Grenoble cedex, France}
\par
\normalsize
\end{center}
\vspace{0.5cm}
\begin{center}
\parbox{16cm}{\underline{abstract}: The possible sources of light nuclei populations 
observed recently below the geomagnetic cutoff by the AMS experiment are discussed in terms 
of nuclear processes: fragmentation of the incoming flux of cosmic \he4 on atmospheric 
nuclei, and nuclear coalescence from proton and \he4 induced reactions. Results of 
simulations for $^{2,3}H$ and $^{3,4}\!H\!e$, are presented and compared to the data.}
\end{center}
\setcounter{page}{1}
The study of particle populations in the earth environment has a long and rich history 
covering the last few decades (see \cite{HIS} for a general overiew of the subject, and the 
references in \cite{PROT1,LEPT} for details). After a period of waning activity, the topics 
is likely to regain interest with the occurence of a new generation of balloon and satellite 
experiments, wich open prospects for data samples of unmatched quantity and quality. This is 
well illustrated with the new data obtained recently in orbit close to earth by the 
precursor flight of the AMS experiment, which achieved on its orbit, a set of high accuracy 
measurements of charged particle flux over a wide range of latitude. 
New features of the proton \cite{PROT1,PROT2} and lepton \cite{LEPT} flux have been 
uncovered, and surprisingly large values of flux measured below the geomagnetic cutoff (GC). 
These features have been successfully interpreted in previous works by the authors 
\cite{PAP1,PAP2}, referred to as I in the following. In addition, small but significant 
populations of \het and deuterium ($D$ or $^2H$) particles were also measured below GC, 
with kinetic energies extending beyond 1~GeV per nucleon \cite{HELI,DEUT}. 
This paper completes a set of three reports devoted to the interpretation of the new AMS 
data. Its purpose is to investigate the possible origins of the measured \het and $D$ flux 
using the same approach as reported in I. 
\par
Since particles below GC cannot be primary cosmic rays (CR), they have to be produced by 
nuclear reactions between incoming CRs (mainly $p$ and \he4), and atmospheric nuclei 
(mainly $^{14}N$ and $^{16}O$). The pattern of Z=2 particle spectra observed above and below 
GC is highly peculiar since only \he4 are observed above GC whereas only \het are found 
below GC, with however a small admixture of the other isotope being compatible with the data 
in both cases \cite{HELI}. This pattern, together with the relative population of light 
nuclei, provide clues to the dynamical origin of the subGC particles (or Albedo particles 
in the geophysical terminology). It can be understood qualitatively and evaluated 
quantitatively, in terms of the nuclear reaction mechanisms involved in the production 
process. 
\par
Collisions between nuclear systems such as those of interest here, for incident energies 
beyond the nuclear fermi energy ($\approx$~35~MeV/nucleon) region, have very characteristic 
distributions. The rapidity distributions of reaction products at forward angles display 
two peaks centered around projectile and target rapidities, corresponding to projectile and 
target fragmentation in peripheral collisions respectively, and an intermediate plateau 
between these limits stemming from more central collisions (see for example \cite{BE95}).  
Experimentally, the (projectile) fragmentation regime sets in as low as 20~MeV kinetic 
energy per nucleon \cite{BU76}. The features of the measured differential cross sections at 
small angles can be accounted for in the fragmentation model \cite{GO74}, with the width of 
the fragmentation peak being governed by nuclear motion and nucleon arrangement probability. 
\par
With the increasing production angles, the target-like and projectile-like peaks move to the 
mid-rapidity region, leaving a single broad mid-rapidity peak surviving at the largest 
angles \cite{BE95}. This latter kinematical domain is associated to smaller impact 
parameters and larger energy dissipation due to larger density overlaps in the collisions. 
In this region, the simple fragmentation picture \cite{GO74} fails and the description of 
the collision in terms of the various models based on thermodynamics and 
spectator-participant pictures \cite{WE76,GO77,MO91} are more appropriate  
The energy distributions of very light fragments like $^{1,2,3}H$, $^{3,4}H\!e$, produced at 
large scattering angles in such collisions, can be described by means of a variety of models 
which are all variations approximating more or less successfully the complex multiple 
nucleon-nucleon scattering processes and nuclear collective effects governing these 
collisions \cite{WE76,GU81,BO82} (see also \cite{GO77,MO91,NA81}). 
In this context, the coalescence model \cite{BU63,LL95} stands as the most successful 
phenomenological approach of the production cross section for light nuclei, with a 
remarkable ability to reproducing data over a very wide range of incident energies 
\cite{MO91,NA81,GA85,AB87}. In this approach, nucleons coalesce into clusters whenever they 
fall within the coalescence radius(momentum) in the final state of the collision. Note that 
the models used here are considered from a purely practical point of view, and not be 
discussed in their foundations. 
\par
For the present purpose, the phenomenological perspective can be summarized the following 
way: Small production angles are dominated by velocity-conserving projectile-like 
fragmentation products, while at large angles the coalescence production of light fragments 
of much lower energy per nucleon, dominates the cross section. In the fragmentation picture,
the cross-section is expected to decrease with the decreasing mass number of the (projectile
-like) fragment (larger probability for smaller number of nucleon transferred \cite{BU76,
MO81}), whereas conversely, in the coalescence model it is decreasing with the increasing 
fragment mass number (larger probability for smaller cluster mass). 
In the case of \he4 projectile, projectile-like particles are also light fragments, then 
likely to be produced either by fragmentation or by coalescence, wheareas obviously, protons 
can only induce production of coalescence fragments.
\par
Projectile fragmentation in heavy ion collisions has been extensively studied experimentally 
in the past \cite{GO78}. A few \he4 fragmentation studies have been reported in the 
literature over the momentum range of interest here \cite{BI77,AN83,AR87}. The production 
of non velocity-conserving light fragments in p and \he4 induced collisions is also 
fairly well documented experimentally \cite{GO77,NA81,MO91}. 
\par
The \het spectrum observed in the AMS experiment \cite{HELI} between the low energy cutoff of 
the spectrometer and up to about 1~GeV/nucleon (see figures \ref{HESPECT} and \ref{HE3} below), 
could originate from the two types of reactions described above:
\par\noindent
1) It could be produced by CR \he4 particle fragmentation or fragmentation-like process on 
atmospheric nuclei (this assumption was also quoted recently in \cite{LI01}). 
The quasi absence of subGC \he4 is easy to understand in this framework since secondary 
\he4 from fragmentation could only originate from the very small CR flux of heavier nuclei, 
i.e., mostly $^{12}C$ and $^{16}O$ (the possible \he4 yield from nuclear evaporation being 
expected at energies below the AMS sensitivity range). 
However, the basic velocity-conserving property discussed above, is clearly not met here 
since incident CR \he4 have momenta larger than about 6~GeV per nucleon in the equatorial 
region, while detected {\het} and $D$ particles have less than 1~GeV per nucleon (see 
details below). 
\par\noindent
2) The observed \het population could also be produced by nuclear coalescence \cite{BU63} 
(see \cite{MO91} for a review of models) from $p$ and \he4 induced collisions on atmospheric 
nuclei \cite{NA81,BO88,SA88}. It would explain as well the absence of subGC \he4 as discussed 
above, 
with a larger population expected below GC for \het than for \he4 particles, by typically one 
order of magnitude \cite{NA81}, which is compatible with the experimental observation. 
\par
The experimental situation then favors the coalescence assumption. The measured flux however, 
depend on the particular dynamics of each production reaction and of the acceptance of the 
magnetosphere to the particle considered, and only a detailed investigation can provide 
a definite answer.
\par
The inclusive spectrum of light nuclei flux at the altitude of AMS (390-400km) have been 
calculated by means of an evolved version of the same simulation program as described in I. 
CR protons and Helium 4 are generated with their natural abundance and momentum 
distributions. They are propagated inside the earth magnetic field using 4th order 
adaptative Runge-Kutta integration of the equation of motion. They are allowed to interact 
with atmospheric nuclei and to produce secondary nucleons $p$,$n$, $^{2,3}\!H$, and 
$^{3,4}\!He$, particles with cross sections and multiplicities as discussed below. Each 
secondary particle is then propagated and allowed to interact as in the previous step. Only 
destructive interaction is taken into account for light nuclei, except \he4. A reaction 
cascade can thus develop through the atmosphere. The reaction products are counted when they 
cross the virtual sphere at the altitude of the AMS spectrometer, upward and downward. All 
charged particles undergo energy loss by ionisation. Each event is propagated until the 
particle disappears by nuclear collision, stopping in the atmosphere by energy loss, or 
escaping to outer space. See I for details.
The small production cross section, and then multiplicity, combined with the small 
magnetosphere acceptance, for the particles of interest here, require a tremendous number 
of events to be generated, and then a huge computer time, for significant statistics to be 
reached. This difficulty has been turned around by enhancing numerically the production 
multiplicity, with the produced events being weighted by the inverse enhancement factor. It 
has been checked carefully that no distortion of the physics involved could be induced by 
using this method, in the studied case.
\par 
The CR proton and helium flux used in the calculations were those measured by AMS 
\cite{PROT1,PROT2,HELI}. The values of the total $p$ and \he4 reaction cross sections used 
were based on the parametrization of \cite{LE83}, the latter being checked on the 
measurements performed on carbon target \cite{JA78}. The production cross sections for light 
nuclei have been implemented and run simultaneously in the event generator on the basis of 
the two models described above. 
\par\noindent
A) - For the \he4 fragmentation cross section, the model from \cite{GO74} was used. 
In this model, the fragment production cross section is proportional to 
$e^{-\frac{P^2}{\sigma^2}}$ in the projectile reference frame, with 
$\sigma^2=\sigma_0^2\frac{A_f(A_p-A_f)}{A_p-1}$, and $P,A_p,A_f$ being the fragment momentum in 
this frame, projectile mass number, and fragment mass, respectively.
The value of Fermi momentum related parameter $\sigma_0$=100~MeV/c was chosen in agreement 
with the results of \cite{GO74,MO81}. The total \he4 fragmentation cross-sections into \het, 
$^3H$, and $D$ used in the program, were borrowed from measurements on Carbon target 
\cite{AB81}, and corrected for their $A^{1/3}$ dependence. 
\par\noindent
B) - In the coalescence model, the invariant differential production cross sections 
for composite fragments with mass $A$, are related to the nucleon production cross section 
by a simple power law:
\begin{displaymath}
(E_A\frac{d^3N_A}{d\vec{p_A}^3})=B_A\cdot(E_p\frac{d^3N_p}{d\vec{p_p}^3})^A ,
\label{COAL}
\end{displaymath}
with $\vec{p_A}=A\cdot\vec{p_p}$. This relation provides straightforwardly the momentum 
spectrum of mass A fragments as a function of the nucleon spectrum at the same production 
angle. The inputs of the event cross section calculation then consist only of the value of 
the $B_A$ parameter and of the proton production differential cross-section. The coalescence
parameters for $p$ induced collisions have been found to have approximately constant values 
through the energy range from 0.2~GeV/n up to 70~GeV/n \cite{SI95}. The following values, 
averaged from \cite{MO91,NA81,AB87,SA88,SA94}, were used in the program: $B_2=2.5\cdot10^
{-2}$ for deuteron, $B_3=2.5\cdot10^{-4}$ for \het, $B_3'=4\cdot10^{-4}$ for $^3H$, and 
$B_4=4\cdot10^{-6}$ for \he4, in units (GeV$^2$/c$^3$)$^{(A-1)}$. The accuracy on these 
values is estimated to be within $\pm$30\% for $B_2$, and $\pm$50\% for $B_3$ and $B_3'$. 
See \cite{NA81} for $B_4$. The proton induced proton production was generated as described 
in \cite{DE00} using the paramertization from \cite{KALI}, while for incident \he4 
particles, the same spectral shape as for protons was used, with the appropriate scaling to 
take into account the experimental total reaction cross section \cite{JA78} and proton 
production cross section (multiplicity) for \he4 collisions on nuclei \cite{AR87,BA93}.
\par
\begin{figure}[htb]
\begin{center}
\hspace{-2cm}
\epsfysize=16cm
\epsfbox{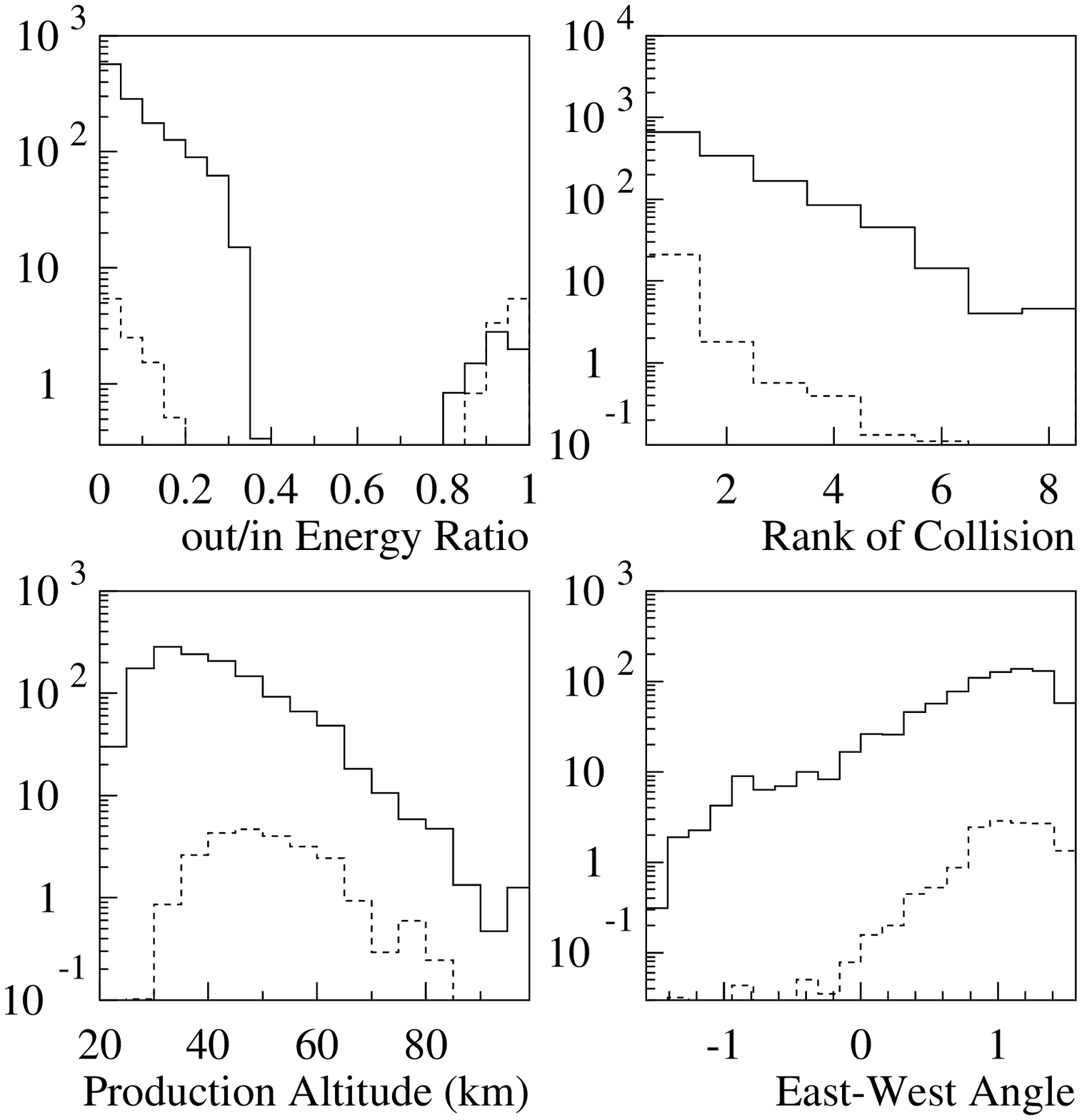} \\
\end{center}
\caption{\small\it Features of the simulated samples for \het (dashed histograms) and $D$ 
particles (solid  histograms). From top to bottom and from left to right: Altitude of 
production (km), rank in the atmospheric cascade, fractional energy of particles, and 
Est-West angle distributions (rad).}
\label{FEAT}
\end{figure}

The simulation has been run for 2 10$^8$ incident primary proton and helium particles 
generated at the injection sphere. This number corresponds to a sampling time of 
4 10$^{-12}$~s of the cosmic ray flux. The results are shown on figures \ref{FEAT} to 
\ref{DEUTONS}. Note that no adjustable parameter was involved in the calculations. 
\par
The general features of the simulated sample are quite similar to those already reported in 
I for protons. In particular, concerning the trapping of particles in the earth magnetic 
field, the same trend is observed for the population to be confined with a long lifetime 
($>10~s$) in the region of equatorial latitudes. The same class of low energy very long 
lifetime but low crossing multiplicity, quasi-polar population as observed in I, is found 
here, likely corresponding to the population of more outer belts than studied here. 

Figure~\ref{FEAT} shows the some basic distributions of the produced \het and 
$D$ particles, crossing the detection altitude (no angular acceptance involved at this 
level). The fractional energy of the produced fragment with respect to the energy of the 
incident article (top left) shows a clear distinction between the coalescence yield at low 
energy and the approximately velocity-conserving fragmentation products at high energy. Also
seen is the strong dominance of the $D$ coalescence yield, larger than the \het yield by 
two orders of magnitude, and the much less different fragmentation yields, due to the 
particular structure of \he4 which fragments into $D$ with a naturally higher multiplicity 
than in the general case. The rank distribution in the atmospheric cascade for \het (top 
right, solid histogram) displays a prominent peak for rank one. This originates from the 
fragmentation \het component which occurs mostly at the first interaction of the primary 
CR \he4. Coalescence particles are seen to be produced up to more than 8 generations of 
collision in the cascade. The mean production altitude (bottom left) predicted for deuterons 
(40~km) in agreement with our previous calculations (see \cite{BKLET} for the experimental 
context) is significantly lower than for (coalescence plus fragmentation) \het (51~km). This 
is also due to the fragmentation process which occurs at the first collision and thus at 
higher altitude on the average (dotted histogram). The \het coalescence yield (not shown) 
has a similar distribution as for $Ds$. The same strong East-West effect (bottom right) 
responsible for the lepton asymmetry reported in \cite{PAP2}, is also seen here as it could 
be expected.
\par
Figure~\ref{HESPECT} shows the comparison between the flux spectra measured by AMS (full 
circle) \cite{HELI} and the simulation results (histograms), taking into account the 
spectrometer acceptance (30 deg), for 3 bins in latitude. The solid 
histograms correspond to the \he4 CR flux above GC. It is seen that they reproduce fairly 
well the experimental distributions, with however a tendancy to underestimate the 
experimental CR flux close to GC. This defect was not observed in our previous works on the 
protons \cite{PAP1} and lepton \cite{PAP2} flux. The dotted histograms correspond to \het 
particles produced by \he4 fragmentation. The expected yield is seen to be significant only 
above GC, and the differential flux appears to be more than two order of magnitude smaller 
than the primary \he4 flux, except for the few bins very close to the cutoff. This value is 
small compared to the known \het CR flux \cite{RE98,BE93,WE91} which ratio to the CR \he4 
flux is about 10\% for this momentum range, i.e. about ten times larger than the value 
calculated here. This result is compatible with the AMS measurements (see figure 4 in 
\cite{HELI}). The dashed and gray-shaded histograms correspond to CR $p$ plus \he4 induced 
coalescence \het and \he4 flux respectively. As expected from the introductory discussion, 
the \het flux is more than one order of magnitude larger than for \he4. These flux are lying 
below GC, and the simulated \het spectra account pretty well for the measured spectra in 
magnitude and shape, with some underestimate of the data however in the higher subGC 
momentum range for the equatorial and intermediate latitudes. 
\par
Upper figure~\ref{HE3} shows the (downwards) \het particle spectrum measured by AMS, in 
kinetic energy per nucleon, integrated over the corrected geomagnetic (CGM, see I for the 
precise definition) 
latitude latitude range $|\theta_M|<0.6$~rad, compared to the simulation results 
(coalescence yield, solid histogram). The calculated values are seen to 
be in agreement with the data within a factor of about 2 over the whole energy domain and 
over a dynamical range covering two orders of magnitude, which can be considered as a very 
good overall agreement. The small \he4 coalescence yield (dashed histogram) predicted is 
expected to be more than one order of magnitude smaller, a value compatible with the AMS 
conclusions in which an experimental upper limit of 10\% for this ratio was set.
Lower figure~\ref{HE3} compares the experimental (full circles) and simulated (histograms)
energy-integrated \het flux below GC as a function of the geomagnetic latitude. The full 
histogram corresponds to the total expected yield. It accounts quite well for the measured 
flux except for the highest latitude bin. The dashed histogram gives the \he4 induced 
contribution, which is seen to amount to 40-50\% of the total yield. This is due to the 
combination of a larger total reaction cross section and larger proton multiplicity (with 
then a larger coalescence probability) with \he4 than with proton incident particles, which 
enhances the \he4 induced coalescence yield.    
\par
Figure~\ref{DEUTONS} shows the expected spectral yields for $^2H$ and $^3H$ particles, 
for a set of bins in latitude. Note that although the two contributions were included, 
these yields are almost exclusively due to coalescence. The results for the $^2H$ spectra 
below GC are in very good agreement in magnitude and in shape with the preliminary AMS 
results \cite{DEUT}.  
This ability of the calculations to successfully and consistently reproducing both the \het 
and $D$ flux at the same altitude of measurement with the same production model constitutes 
a very strong indication that these particles do originate from the same coalescence 
mechanism.
\par
Some $^3He$ flux measurements have been reported recently at lower energies and larger 
distance from earth in the region of the inner belt \cite{WE95}. This flux could not be 
definitely interpreted in the quoted work. It could certainly be investigated in the 
present approach.
\par
In conclusion, it has been shown that the AMS measurements of the \het and $D$ particle flux
can be reproduced consistently and simultaneously, together with the proton secondary flux 
as reported in I, by a simulation incorporating the interactions between Cosmic Ray flux, 
earth magnetosphere and atmosphere, and assuming the \het and $D$ particles to be produced 
by coalescence of nucleons in the CR proton and \he4 induced nuclear collisions with 
atmospheric nuclei. The \he4 fragmentation products appear not to contribute at a detectable
level to the flux measured below the geomagnetic cutoff.
\par
These results also clearly point to the interest for the future satellite experiments to 
have a particle identification capability covering this mass region and extending over a 
large kinetic energy range down to about 0.2~GeV per nucleon, in order to allow the 
experiments to collect data samples larger by more than two orders of magnitudes than those
analyzed here, and then to make possible the achievement of a much more detailed 
investigation of the issues addressed in the present work.
\par
%
%

%

%
\begin{figure}[htb]
\begin{center}
\hspace{-2cm}
\epsfysize=16cm
\epsfbox{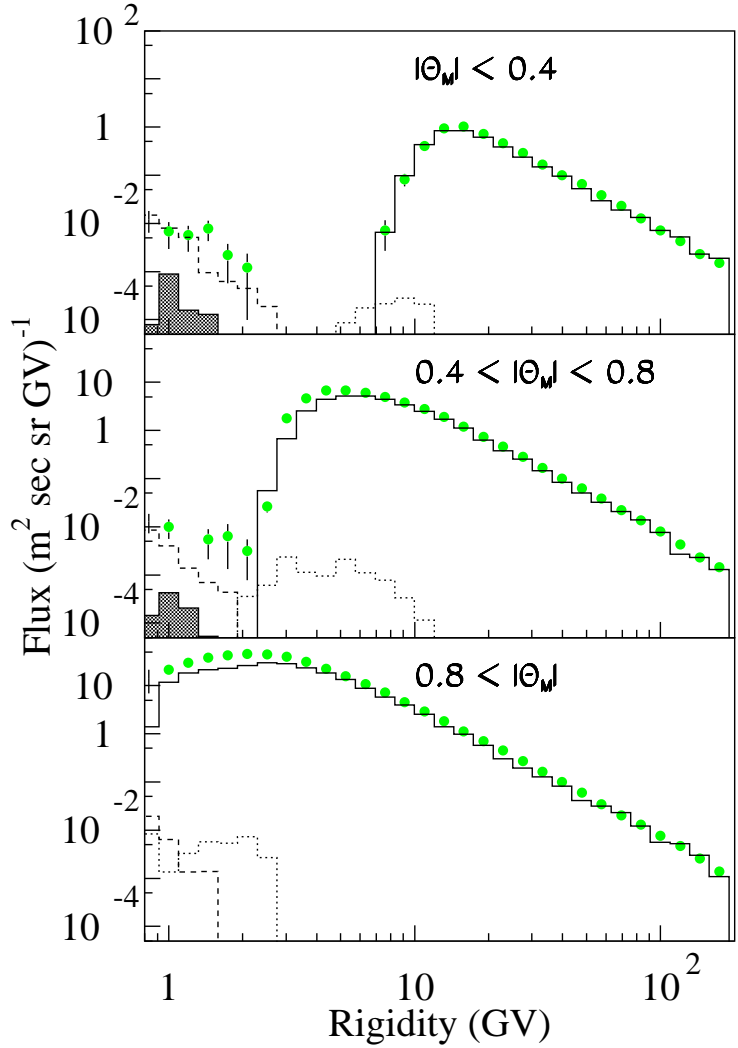} \\
\end{center}
\caption{\small\it Comparison of the Helium spectra measured by AMS (\cite{HELI} full 
circles) at various CGM latitudes $\theta_M$, with the simulation results (histograms). 
Full line : \he4 above GC; Dotted: \het from \he4 fragmentation;  Dashed: coalescence \het; 
shaded: coalescence \he4. See text for details.}
\label{HESPECT}
\end{figure}
%
\begin{figure}[htb]
\begin{center}
\hspace{-2cm}
\epsfysize=16cm
\epsfbox{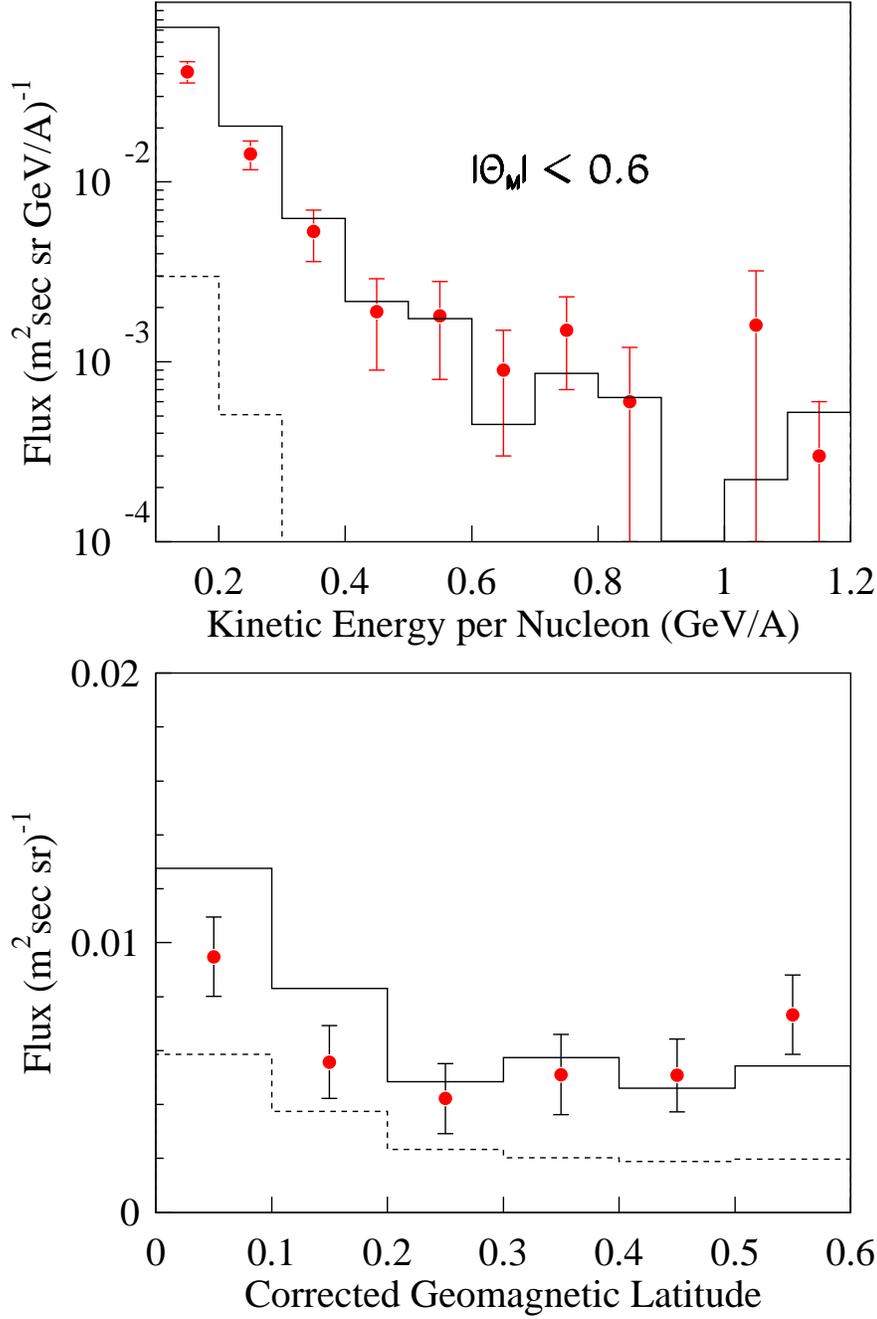} \\
\end{center}
\caption{\small\it Top: Experimental energy spectrum of the subGC \het flux integrated
over $\theta_M<0.6~rad$, measured by AMS \cite{HELI}, compared to simulation results (solid 
histogram). The dashed histogram corresponds to the coalescence \he4 flux. 
Bottom: Energy-integrated \het experimental distribution (E~$<$~1.2~GeV/A) as a function of 
the CGM latitude (full circles) compared to simulation results for proton and \he4 induced 
flux. The dashed histogram shows the \he4 induced contribution.}
\label{HE3}
\end{figure}
%
\begin{figure}[htb]
\begin{center}
\hspace{-2cm}
\epsfysize=16cm
\epsfbox{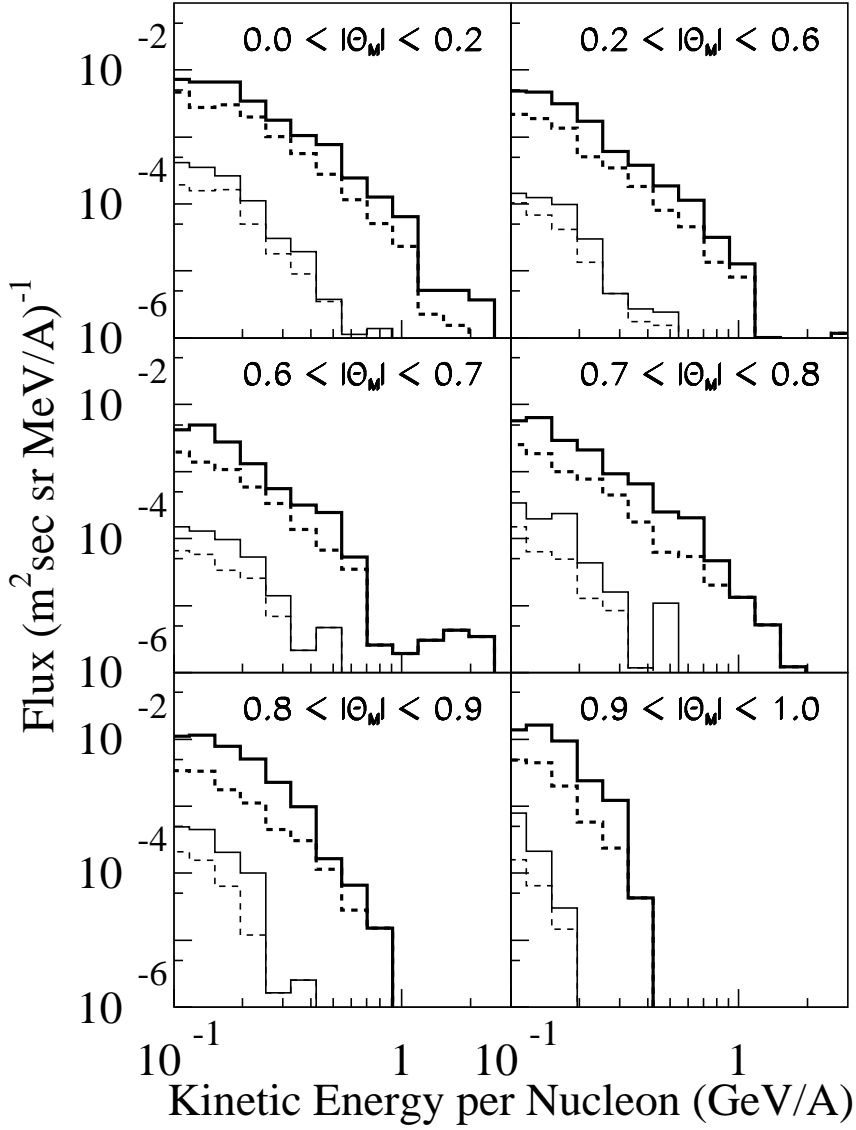} \\
\end{center}
\caption{\small\it Prediction for the subGC flux of deuterium $^2\!H$ (thick histograms) and 
tritium $^3\!H$ particles (thin histograms) for various bins of latitude between equator and 
polar region (CGM bin values $\theta_M$ given in radian). Solid line : full ($p$ plus \he4) 
yield;  Dashed line : \he4 yield.}
\label{DEUTONS}
\end{figure}
\end{document}